\documentclass[12pt]{article} 
\usepackage{graphicx}
\renewcommand{\theequation}{\thesection.\arabic{equation}}
\newcommand{\be}{\begin{equation}}
\newcommand{\ee}{\end{equation}}
\newcommand{\bear}{\begin{eqnarray}}
\newcommand{\eear}{\end{eqnarray}}
\newcommand{\ba}{\begin{array}}
\newcommand{\ea}{\end{array}}
\newcommand{\lae}{\begin{array}{c}\,\sim\vspace{-21pt}\\<
\end{array}}
\newcommand{\gae}{\begin{array}{c}\,\sim\vspace{-21pt}\\>
\end{array}}

 \newcommand{\CQ}{{\cal Q}}
\newcommand{\CU}{{\cal U}} 
\newcommand{\CL}{{\cal L}} 
 
\newcommand{\CH}{{\cal H}} \newcommand{\CM}{{\cal M}}
\newcommand{\CW}{{\cal W}}
\newcommand{\CB}{{\cal B}}

\begin{document}

\pagestyle{empty} \begin{titlepage}
\def\thepage {}        

\title{\Large \bf Universal Extra Dimensions and \\ the Higgs Boson Mass
\\ [1.3cm]}

\author{\normalsize
\bf \hspace*{-.3cm} Thomas Appelquist,
Ho-Ung Yee
 \\ \\ {\small {\it
\vspace*{-5cm}
Department of Physics, Yale University, New
Haven, CT 06520, USA \footnote{e-mail: thomas.appelquist@yale.edu, \, ho-ung.yee@yale.edu}
}}\\
 }

\date{ } \maketitle

\vspace*{-7.9cm}
\noindent \makebox[12.7cm][l]{\small \hspace*{-.2cm}
hep-ph/0211023} {\small YCTP-P10-02 } \\
\makebox[11.8cm][l]{\small \hspace*{-.2cm} November 2, 2002 }
{\small } \\

 \vspace*{10.5cm}

  \begin{abstract}
{\small
We study the combined constraints on the compactification scale $1/R$ and
the Higgs mass $m_H$ in the standard model with one or two universal extra
dimensions. Focusing on precision measurements and employing the
Peskin-Takeuchi $S$ and $T$ parameters, we analyze the allowed region
in the $(m_H, 1/R)$ parameter space consistent with current
experiments. For this purpose, we calculate complete one-loop KK mode
contributions to $S$, $T$, and $U$, and also estimate the contributions from
physics above the cutoff of the higher-dimensional standard model. A compactification scale
$1/R$ as low as $250\,{\rm GeV}$ and significantly extended regions of $m_H$
are found to be consistent with current precision data.
}
\end{abstract}

\vfill \end{titlepage}

\baselineskip=18pt \pagestyle{plain} \setcounter{page}{1}

\section{Introduction} \setcounter{equation}{0}

In models with universal extra dimensions (UED's) in which all the standard model fields propagate,
bounds on the compactification scale, $1/R$, have been estimated from precision experiments to be as low as $\sim\,300\,{\rm GeV}$ \cite{Appelquist:2000nn,Agashe:2001xt}.
This would lead to an exciting phenomenology in the next generation of collider
experiments \cite{Cheng:2002ab,Rizzo:2001sd,Macesanu:2002db,Petriello:2002uu}. Above the compactification scale,
the effective theory becomes a higher dimensional field theory whose equivalent description in 4D consists of
the standard model fields and towers of their KK partners whose interactions are very similar to those in the standard model.
Because the effective theory above the compactification scale (the higher dimensional standard model) breaks down at the scale
$M_s$, where the theory becomes nonperturbative, the towers of KK particles must be cut off at this
scale in an appropriate way. The unknown physics above $M_s$ can be described by operators of higher mass dimension whose
coefficients can be estimated.

To obtain the standard-model chiral fermions from the corresponding extra dimensional fermion fields, the higher dimensional
standard model is compactified on an orbifold
to mod out the unwanted chirality by orbifold boundary conditions. For a single (two) universal extra dimension(s),
this is $S^1/Z_2$ ( $T^2/Z_2$ ) \cite{Ponton:2001hq}.
The interactions involving nonzero KK particles are largely determined by the bulk lagrangian in terms of the higher dimensional standard model,
while the effects from possible terms localized at the orbifold fixed points are relatively volume-suppressed.
The KK particles enter various quantum corrections to give contributions to precision measurements. Studies of
their effects on the precision electroweak measurements in terms of $S$ and $T$ parameters \cite{Appelquist:2000nn}, on the flavor
changing process $b\,\rightarrow\,s\,+\gamma$ \cite{Agashe:2001xt}, and on the anomalous muon magnetic moment \cite{Agashe:2001ra,Appelquist:2001jz}
have shown that these effects are consistent with current precision experiments if $1/R$ is above a few hundred ${\rm GeV}$.
The cosmic relic density of the lightest KK particle as a dark matter candidate is also of the right order of magnitude \cite{Servant:2002aq},
and its direct or indirect detection is within the reach of future experiments \cite{Cheng:2002ej,Hooper:2002gs,Servant:2002hb,Majumdar:2002mw}.

In this paper, we address the effects of the new physics above $1/R$ on the combined constraints for the Higgs mass and $1/R$.
Current knowledge of the Higgs mass has been inferred from its contributions to the electroweak precision observables.
Because the new physics in terms of KK partners and higher dimension operators representing physics above $M_s$ also contributes to these
observables, the constraints on the Higgs mass can be significantly altered in the UED framework. 
The effects on the precision observables from non zero KK modes depend on both
$1/R$ and the Higgs mass, $m_H$ (through KK Higgs particles), while the standard model (the zero modes) contributions are functions of $m_H$ alone. We therefore
analyze the allowed region in the $(m_H,1/R)$ parameter space consistent with the current precision measurements.

Current precision electroweak experiments are sensitive to new-physics corrections to fermion-gauge boson vertices and
gauge-boson propagators. The most sensitive fermion-gauge boson vertex is the $Zb\bar{b}$ vertex.
Contributions to it were analyzed in Ref.\cite{Appelquist:2000nn}. The dominant contribution comes from
loops with KK top-bottom doublets:
\be
\delta g_{L}^{b} \sim \frac{\alpha}{4\pi} \frac{m_t^2}{M_j^2}\quad,  
\label{vertex}
\ee
where $M_j\,=\,\sqrt{j_1^2+\cdots+j_\delta^2}/R$, and $j=(j_1,\cdots,j_\delta)$ is a set of indices of KK levels in $\delta$ extra dimensions.
It was noted there that these corrections are less important than the Peskin-Takeuchi $S$ and $T$ parameters \cite{Peskin:1990zt} in constraining UED theories
for the phenomenologically interesting region of $1/R \gg m_t$.
We therefore focus on the Peskin-Takeuchi parameters. 

We consider two possibilities; a single universal extra dimension on $S^1/Z_2$ and two universal extra dimensions on $T^2/Z_2$.
In the case of a single extra dimension, the cutoff effects from physics above $M_s$ are estimated to be negligible and we can do a reliable
calculation of the contributions from KK modes alone. This is not the case for the model with two extra dimensions.
The UED theory on $T^2/Z_2$ is a particularly interesting model because it points to 
three generations \cite{Dobrescu:2001ae} (See also \cite{Fabbrichesi:2001fx}),
and can explain the longevity of protons \cite{Appelquist:2001mj}. 
The neutrino oscillation data can also be accomodated within this model \cite{Appelquist:2002ft}.
However, the sums over the KK particle contributions to precision observables are logarithmically divergent with two extra dimensions,
and effects from above the cutoff $M_s$
must be included. We estimate these effects using higher dimension operators, which makes the analysis only qualitative,
but we can still extract useful information from the results.

In the next section, we describe the calculation of $S,T$ and $U$ from one-loop diagrams with KK particles, and a subtlety involved in this calculation.
In section 3, we estimate the contributions to $S$ and $T$ from physics above the cutoff $M_s$. Sections 4 and 5 are devoted to the details of the analysis
with both a single extra dimension on $S^1/Z_2$ and two extra dimensions on $T^2/Z_2$. We summarize and conclude in section 6.

\section{KK-mode contributions to the $S,T$ and $U$ parameters} \setcounter{equation}{0}


In the analysis of Ref.\cite{Appelquist:2000nn}, it was argued that
the dominant contributions to $S$ and $T$ come from KK modes of the top-bottom quark doublet:
\be
T^t_j\,\sim\,\frac{1}{\alpha}\frac{3m_t^2}{8\pi^2 v^2}\frac{2}{3}\frac{m_t^2}{M_j^2}\quad,\qquad
S^t_j\,\sim \,\frac{1}{6\pi}\frac{m_t^2}{M_j^2}\quad.
\ee
It was shown that the constraint from $T$ is stronger than that from $S$.
The $U$ parameter is numerically much smaller than $S$ and $T$, thus much less important in constraining UED theories.
An important premise in Ref.\cite{Appelquist:2000nn} was that the Higgs mass, $m_H$, is lighter than $250\,{\rm GeV}$.

If the Higgs mass $m_H$ is large, however, the contributions from the standard model Higgs and its higher KK modes become important
and eventually dominate over the KK quark contributions. 
A key point is that the Higgs contribution to $T$ is negative, which
is opposite to the KK quark contribution. (For $S$, both KK quarks and KK Higgs contributions are positive.)
Thus, the two contributions can compensate each other to relax the $T$ constraints, allowing
an extended region in the $(m_H,1/R)$ parameter space. Moreover, a large $m_H$ can also bring important constraints from $S$, requiring
a combined $S$ and $T$ analysis rather than separate ones.
It is thus important to do a more complete analysis allowing for the possibility of a large Higgs mass.

We calculate complete one-loop corrections from a given $j$th KK level of the standard model fields (with a single Higgs doublet)
to gauge-boson self energies:
$\Pi^j_{WW}, \Pi^j_{ZZ}, \Pi^j_{\gamma \gamma}$ and $\Pi^j_{Z \gamma}$ (See the appendix). Here $j$ represents a positive
integer for one extra dimension or a set of $\delta$ non negative integers in the case of $\delta$ extra dimensions.
The total contribution from extra dimensions will be the sum over $j$.
In the large KK mass limit $M_j \gg m_t,m_H$, the
contributions to $S,T$ and $U$ parameters are proportional to $\frac{m_t^2
}{M_j^2},\,{\rm or}\,\frac{m_H^2}{M_j^2}$.  In one extra dimension, there is one KK mode for each positive interger $j$, and the sum converges.
However, in two or more extra dimensions, there are degenerate KK modes having the same $M_j$, which makes the sum divergent. 
With two extra dimensions, the cutoff sensitivity is logarithmic.
In our calculation of $S,T$ and $U$, we use the tree-level formula for the masses of KK particles neglecting
corrections from one-loop gauge interactions and boundary terms localized at the orbifold fixed points \cite{Cheng:2002iz}.
This is justified because these are of one-loop order and the shifts due to them, which are already of one-loop order, are two-loop effects.

Before presenting our results, we discuss a subtlety in the calculation.
The conventional definition of the T parameter is 
\be
\alpha(m_Z)\,T\,\,\equiv\,\,\frac{\Pi_{WW}(p^2=0)}{m_W^2}-\frac{\Pi_{ZZ}(p^2=0)}{m_Z^2}\quad,
\label{defT} 
\ee 
where the $\Pi$ functions are the gauge-boson self energies
arising from new, non-standard-model physics, and $\alpha(m_Z)\approx 1/128$. When the non-standard-model
physics is "oblique" (entering dominantly through the gauge-boson self
energies), this definition corresponds directly to a physical measurement.
An example is provided by a loop of KK modes of the standard-model fermions
such as the top quark. In general, the new physics can also contribute
through vertex corrections and box diagrams. An example of this is provided
by one-loop corrections involving KK gauge bosons. All the pieces must then be
combined to insure a finite and gauge invariant (physical) result. Indeed,
we find in our calculation that the one-loop divergences in $T$ as defined above, arising from KK gauge bosons and KK Higgs bosons, do not
cancel, although $S$ and $U$, as conventionally defined, turn out to be
finite and well defined.

The one-loop contributions to $\Pi_{WW}$ and $\Pi_{ZZ}$ are listed in the appendix. The computation has been done in Feynman gauge.
From the tabulation, one can see that in this gauge, $T$ is UV-divergent at the one-loop level, and that the divergence 
arises from graphs involving loops of KK gauge bosons and KK Higgs bosons.
This indicates that there should be a non-vanishing counterterm for the $T$ parameter of (\ref{defT}).
It can be shown that, because of the constraints from gauge symmetry, 
the counterterm for $T$ is determined by the $A_\mu Z^\mu$-counterterm at the one-loop level.
Once we fix the $A_\mu Z^\mu$-counterterm, corresponding to photon-$Z$ mass mixing, cancelling $\Pi_{Z\gamma}(0)$ to ensure a massless photon propagator,
the counterterm for $T$ is completely determined in terms of $\Pi_{Z\gamma}(0)$. 

As a result, in the basis in which the photon-$Z$ mass matrix is diagonal through one-loop, it can be shown that the modified $T$
parameter including the counterterm takes the form
\be
\alpha(m_Z)\,\tilde{T}\,\,\equiv\,\,\frac{\Pi_{WW}(0)}{m_W^2}-\frac{\Pi_{ZZ}(0)}{m_Z^2}
-2\,{\rm cos}\theta_w\,{\rm sin}\theta_w\,\frac{\Pi_{Z\gamma}(0)}{m_W^2}\quad.
\label{defTnew} 
\ee 
It can be checked explicitly from the appendix that this expression is UV-finite. As the finiteness originates from
a certain relation between counterterms determined by gauge symmetry, it is true in any gauge.
Of course, $\tilde{T}$ is not, in general, a gauge-invariant, physical observable unless it is combined with vertex corrections and box diagrams.

The important observation, however, is that the contribution of the 
Higgs-boson KK modes
to the vertices and box diagrams are negligible at the one-loop level since 
they are
suppressed by small Yukawa couplings when they
couple to the light external fermions. Thus the dominant contributions to 
$\tilde{T}$
when $m_H$ and $m_t$ are large compared to the gauge-boson masses, must by
themselves be gauge invariant. It is straightforward to determine these from the appendix. For a $j$th KK level,
\bear
\tilde{T}^j_{\rm KK Higgs}\approx\frac{1}{4\pi}\frac{1}{c_w^2}
\,f_T^{\rm KK Higgs}\bigg(\frac{m_H^2}{M_j^2}\bigg)
\quad, \qquad
\tilde{T}^j_{\rm \,\,KK \,top\,\,}\approx\frac{1}{\alpha(m_Z)}\frac{3\,m_t^2}{8\pi^2v^2}
\,f_T^{\rm \,\,KK \,top\,\,}\bigg(\frac{m_t^2}{M_j^2}\bigg)
\quad,
\label{KKhigtop}
\eear
where $v=246\,{\rm GeV}$ is the VEV of the zero mode Higgs boson and
\bear
f_T^{\rm KK Higgs}(z)&=&\frac{5}{8}-\frac{1}{4z}+\bigg(-\frac{3}{4}-\frac{1}{2z}+\frac{1}{4z^2}\bigg)\log(1+z) \quad , \nonumber \\
f_T^{\rm \,\,KK \,top\,\,}(z)&=&1-\frac{2}{z}+\frac{2}{z^2}\log(1+z)\quad . 
\label{ftnsforT}
\eear
Note that the KK Higgs boson contributions to $\tilde{T}$ are negative, while the contributions from the KK top quarks are positive. 

\begin{figure}[t]
\begin{center}
\scalebox{1}[1]{\includegraphics{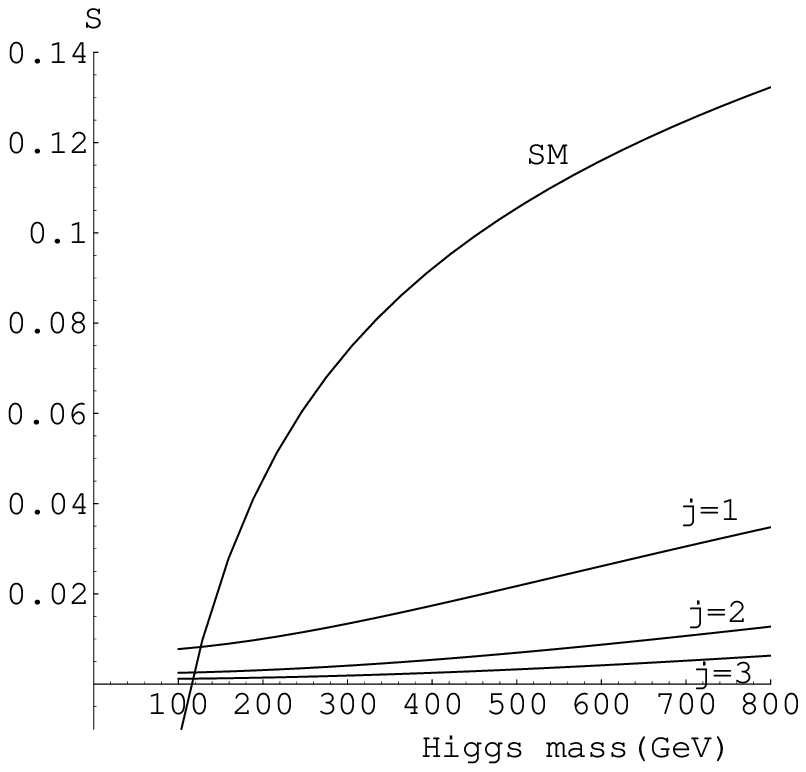}\includegraphics{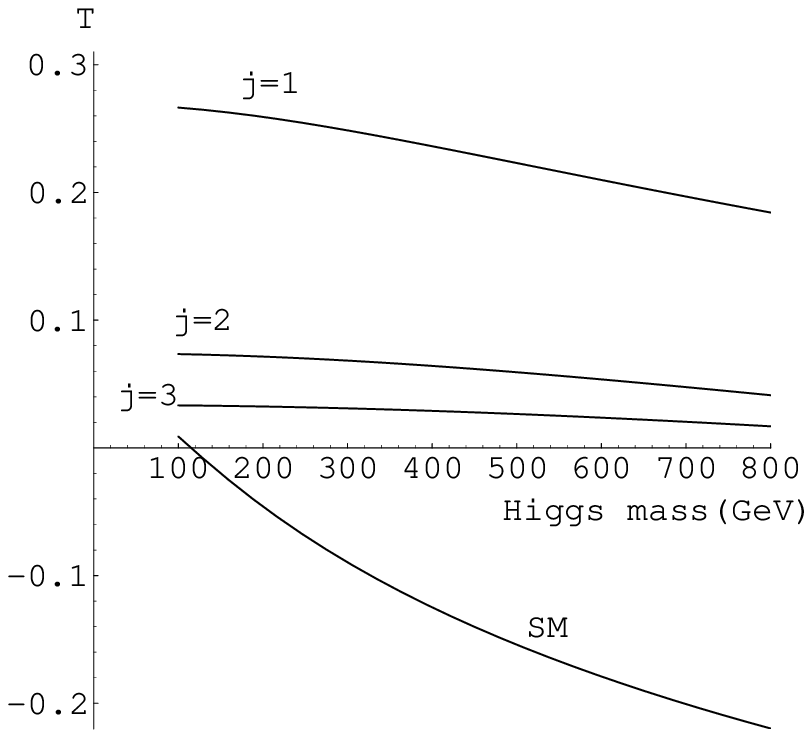}}
\par
\vskip-2.0cm{}
\end{center}
\caption{\small Contributions to $S$ and $T$ from the standard model(the zero modes) and $j$'th KK levels of one UED compactified on $S^1/Z_2$.
Here, $1/R=400 \,{\rm GeV}\,,\,m_{H}^{\rm ref}=115\,{\rm GeV}$ and $m_t=174\,{\rm GeV}$.}
\label{kkmodes}
\end{figure}
By contrast, the typical size of one-loop KK gauge-boson contributions to $\alpha\tilde{T}$, vertex corrections, or box diagrams is of order
\be
\frac{\alpha}{4\pi}\,\frac{m_W^2}{M_j^2}\quad.
\ee
Clearly these are negligible compared to the contributions (\ref{KKhigtop}) when $m_H^2,\,m_t^2\,\gg\,m_W^2$.
This leads us, to a good approximation, to neglect them
and to focus on the dominant, gauge-invariant, oblique contributions (\ref{KKhigtop}) in our numerical analysis.
 
By the same reasoning, the (gauge-dependent) KK gauge-boson contributions to $S$ and $U$ can be neglected. From the appendix, one can write down the
dominant, gauge-invariant expressions for $S$, which are similar to (\ref{KKhigtop}), arising from KK Higgs bosons and KK top quarks:
\bear
S^j_{\rm KK Higgs}\approx\frac{1}{4\pi}
\,f_S^{\rm KK Higgs}\bigg(\frac{m_H^2}{M_j^2}\bigg)
\quad, \qquad
S^j_{\rm \,\,KK \,top\,\,}\approx\frac{1}{4\pi}
\,f_S^{\rm \,\,KK \,top\,\,}\bigg(\frac{m_t^2}{M_j^2}\bigg)
\quad, 
\label{KKhigtopS}
\eear
where
\bear
f_S^{\rm KK Higgs}(z)&=&-\frac{5}{18}+\frac{2}{3z}+\frac{2}{3z^2}+\bigg(\,\frac{1}{3}-\frac{1}{z^2}-\frac{2}{3z^3}\bigg)\log(1+z) \quad , \nonumber \\
f_S^{\rm \,\,KK \,top\,\,}(z)&=&\frac{2z}{1+z}-\frac{4}{3}\,\log(1+z) \quad . 
\label{ftnsforS}
\eear
These, together with (\ref{KKhigtop}), are the basis of our numerical calculations.


In Fig.\ref{kkmodes}, we show contributions to $S$ and $T$ from different $j$th levels in terms of the Higgs mass, for a
representative value of $1/R$, in the case of a single extra dimension on $S^1/Z_2$. 
We also include the standard model contributions from the Higgs (zero mode) after fixing
the reference Higgs mass at $115\, {\rm GeV}$. The contributions from higher KK levels become small rapidly,
consistent with the decoupling behavior \cite{Appelquist:tg}. 
Results for the case of two extra dimensions on $T^2/Z_2$
exhibit similar behavior for each $j$th level, though we must take into account degeneracy when summing them.

\section{Contributions to $S$ and $T$ from physics above $M_s$}

Because our effective theory breaks down at the cutoff scale $M_s$,
we also estimate the contributions from physics above this scale by examining the relevant
local operators of higher mass dimension, whose coefficients incorporate unknown physics above $M_s$. To find the operators that give
direct tree-level contributions to $S$ and $T$, it is convenient to use the matrix notation for the Higgs fields,
\be
\CM\equiv \left(\,i\sigma^2 \CH^*\,,\,\CH\,\right)=\left(\ba{ccc}h^{0*} & h^+ \\ -h^{+*} & h^0 \ea \right).
\ee
Here, the $\CM$, $\CH$ and all calligraphic fields in the following are the fields in $(4+\delta)$ dimensions, whereas the
corresponding roman letters will represent the 4 dimensional zero modes after KK decomposition.
The $SU(2)_L \times U(1)_Y$ gauge rotation is $\CM\,\rightarrow\,U_L(x)\CM e^{-i\alpha (x)\sigma^3}$ and the covariant derivative is
\be
D_{\alpha}\CM\,=\,\partial_{\alpha}\CM\,+\,i\,\hat{g}\,\CW_{\alpha}^{\,a}\,\frac{\sigma^a}{2}\,\CM\,-\,i\,\hat{g}'\,\CB_{\alpha}\,\CM\,\frac{\sigma^3}{2}\quad,
\ee
where $\CW_{\alpha}^{\,a},\,\CB_{\alpha}$ are the gauge fields in $(4+\delta)$ dimensions and $\hat{g},\,\hat{g}'$
are the corresponding $(4+\delta)$-dimensional gauge couplings whose mass dimension is $-\frac{\delta}{2}$. The mass dimension of
$\CW_{\alpha}^{\,a},\,\CB_{\alpha}$ and $\CM \,(\,\CH)$ is $(1+\frac{\delta}{2})$.
The gauge invariance
dictates that the Higgs potential up to quartic order (i.e. up to mass dimension $(4+2\delta)$ ) depends only on $\frac{1}{2}{\rm Tr}[\CM^\dagger\CM]=\CH^\dagger\CH$,
which implies the enlarged $SU(2)_L \times SU(2)_R$ symmetry,
\be
\CM\,\rightarrow\,U_L\CM U_R \quad.
\ee
After the zero mode Higgs field gets a VEV, $<M >=\frac{v}{\sqrt{2}}\,\mathbf{1}$, $v=246\,{\rm GeV}$, this symmetry is broken down to the diagonal custodial
$SU(2)_C$ which protects $T$ at tree level. Hypercharge interactions violate custodial $SU(2)_C$, inducing
nonzero $T$ at loop level.

When we consider operators of higher mass dimension, however, the gauge invariance can no longer prevent operators that violate
the custodial symmetry. There is one independent, custodial symmetry-violating operator of the lowest mass dimension\footnote{Other possible operators can be shown to be equivalent to (\ref{Toperator}) up to additive custodial-symmetric operators.}
$(6+2\delta)$:
\bear
&&c_1\cdot\frac{\hat{\lambda}}{2^2\cdot2!\cdot M_s^2}\,{\rm Tr}[\sigma^3(D_\alpha\CM)^\dagger\CM]\cdot{\rm Tr}[\sigma^3(D^\alpha\CM)^\dagger\CM]\quad, \nonumber \\
&&=c_1\cdot\frac{\hat{\lambda}}{2^2\cdot2!\cdot M_s^2}\,
\big(\CH^\dagger \stackrel{\leftrightarrow}{D_\alpha}\CH\big)\big(\CH^\dagger \stackrel{\leftrightarrow}{D^\alpha}\CH\big)\quad,
\label{Toperator}
\eear
where $\alpha=1,\dots,(4+\delta)$. We have extracted the $(4+\delta)$-dimensional Higgs self coupling $\hat{\lambda}$, of mass dimension $-\delta$,
which appears in the quartic interaction between four Higgs fields,
\be
\CL_{(4+\delta)}\quad\supset\quad\frac{\hat{\lambda}}{2!}\big(\CH^\dagger\CH)^2=\frac{\hat{\lambda}}{2^2\cdot 2!}\big({\rm Tr}[\CM^\dagger\CM]\big)^2\quad,
\label{quartic}
\ee
expecting that $\hat{\lambda}$ reflects the strength of the underlying dynamics responsible for
similar kinds of four-Higgs interactions. Except for the custodial symmetry violation, the operator (\ref{Toperator}) simply has two more
derivatives than (\ref{quartic}) and we have pulled out all the expected factors (including various numerical counting factors)
in writing (\ref{Toperator}). We then expect that
$c_1$ should be a constant no larger than of order unity. If there is a suppression of the custodial symmetry violation,
$c_1$ will be small compared to unity. 

After KK decomposition, the relevant 4D operator from (\ref{Toperator}) is obtained after replacing $(4+\delta)$-dimensional fields with
the corresponding 4D zero modes,
\be
\CM\,(\,\CH\,)\,\rightarrow\,\frac{\sqrt{2}}{(2\pi R)^{\delta /2}}\,M\,(\,H\,) \quad,
\label{subs}
\ee
and integrating over the extra $\delta$ dimensions, $\int d^{\delta}y=(2\pi R)^{\delta}/2$. (The factor 2 is from the $Z_2$ orbifold).
Also replacing $\hat{\lambda}$ with the 4D Higgs self coupling $\lambda$,
\be
\hat{\lambda}=\frac{(2\pi R)^{\delta}}{2}\,\lambda\quad,
\ee
the resulting 4D operator is
\bear
&&c_1\cdot\frac{\lambda}{2^2\cdot2!\cdot M_s^2}\,{\rm Tr}[\sigma^3(D_\mu M)^\dagger M]\cdot{\rm Tr}[\sigma^3(D^\mu M)^\dagger M] \quad,\nonumber \\
&&=c_1\cdot\frac{\lambda}{2^2\cdot2!\cdot M_s^2}\,
\big(H^\dagger \stackrel{\leftrightarrow}{D_\mu}H\big)\big(H^\dagger \stackrel{\leftrightarrow}{D^\mu}H\big)\quad .
\label{4dToperator}
\eear
Note that $(2\pi R)$ factors have dropped out in the expression (\ref{4dToperator}).
The contribution to $T$ from physics above $M_s$ can be estimated from (\ref{4dToperator}) to be
\be
T^{UV}=c_1\cdot\frac{\lambda}{2^2\cdot2!\cdot M_s^2}\cdot\frac{2\,v^2}{\alpha(m_Z)}=c_1\cdot\frac{\lambda\,v^2}{4M_s^2\cdot\alpha(m_Z)}=
c_1\cdot\frac{m_H^2}{4M_s^2\cdot\alpha(m_Z)} \quad .
\label{deltaT}
\ee
This result will be used in later sections when we estimate all the contributions to $T$ in one or two extra dimensions.

We next discuss the $S$ parameter. From the definition of $S$,
\be
S=-\frac{16\pi}{g\cdot g'}\,\frac{d}{dq^2}\Pi_{3Y}(q^2)\bigg|_{q^2=0}\quad,
\ee
where $g$ and $g'$ are the 4D gauge coupling constants of $SU(2)_L\times U(1)_Y$,
it is clear that we need an operator that couples the $SU(2)_L$ gauge field $\CW^3$ with
the hypercharge gauge field $\CB$ to describe the contributions to $S$ from physics above $M_s$.
It is not difficult to find the operator with the lowest mass dimension \cite{Appelquist:1993ka},
\be
-c_2\cdot\frac{\hat{g}\,\hat{g}'}{2^2\cdot 2!\cdot M_s^2}\,\,\CB_{\alpha\beta}\,\,{\rm Tr}\big[\,\CM\sigma^3\CM^\dagger\,\CW^{\alpha\beta}\,\big]\quad .
\label{Soperator}
\ee
For each field strength, $\CW_{\alpha\beta}$ and $\CB_{\alpha\beta}$, we have included a counting factor of $\frac{1}{2}$.
We have also pulled out the $(4+\delta)$-dimensional
gauge couplings, $\hat{g}$ and $\hat{g}'$, expecting that the $\CW$ and $\CB$ fields naturally couple to the underlying dynamics
that generates (\ref{Soperator}) with the strength of gauge couplings.
Having done this, we expect $c_2$ to be a constant of order unity.
The corresponding 4D operator from (\ref{Soperator}), after the substitutions (\ref{subs}) and
\be
\big(\CW_{\alpha\beta},\,\CB_{\alpha\beta}\big)\,\rightarrow\,\frac{\sqrt{2}}{(2\pi R)^{\delta /2}}\,\big(W_{\mu\nu},\,B_{\mu\nu}\big)
\quad,\qquad \big(\hat{g},\,\hat{g}'\big)\,=\,\frac{(2\pi R)^{\delta /2}}{\sqrt{2}}\,\big(g,\,g'\big)\quad,
\ee
and the volume integration $\int d^{\delta}y=(2\pi R)^{\delta}/2$, is
\be
-c_2\cdot\frac{g\,g'}{2^2\cdot 2!\cdot M_s^2}\,\,B_{\mu\nu}\,\,{\rm Tr}\big[\,M\sigma^3M^\dagger\,W^{\mu\nu}\,\big]\quad.
\label{4dSoperator}
\ee
This gives the following estimate of $S$ from physics above the cutoff scale:
\be
S^{UV}=c_2\cdot\frac{2\pi\,v^2}{M_s^2}\quad .
\label{deltaS}
\ee
Note again that the final result doesn't depend explicitly on the number of extra dimensions nor the compactification scale, $1/R$.
This estimate will also be useful later when we discuss the case of one or two extra dimensions.

\section{One universal extra dimension on $S^1/Z_2$} \setcounter{equation}{0}

In the case of one extra dimension, the sum over the KK contributions to $S, T$ and $U$ is convergent. Thus we can obtain reliable results
if the convergence is fast enough so that the cutoff effects on the KK sum are insignificant.
We see from Fig.\ref{kkmodes} that the contributions from higher KK levels become small rapidly.
The error of summing only up to the 11'th KK level is estimated to be less than 1\%.
The cutoff $M_s$ is estimated to be $\sim 30\cdot 1/R$ \cite{Appelquist:2000nn}, implying that the cutoff is irrelevant for the KK sum.
Because the standard model also contributes to
the oblique parameters as we change the Higgs mass from $m_{H}^{\rm ref}=115 \,{\rm GeV}$, we must include those in the final $S,T$ and $U$
calculation.

Although the KK sum is insensitive to the cutoff, it is important to check explicitly that
the cutoff effects in terms of higher dimension operators are indeed negligible. From (\ref{deltaT}) and (\ref{deltaS}),
their size can be read conveniently from the following expression:
\bear
&&T^{UV}=c_1\cdot\,1.6\times 10^{-2}\Bigg(\frac{m_H}{200\,{\rm GeV}}\Bigg)^2\Bigg(\frac{300\,{\rm GeV}}{1/R}\Bigg)^2\Bigg(\frac{30}{M_sR}\Bigg)^2\quad,\nonumber \\
&&S^{UV}=c_2\cdot\,4.7\times10^{-3}\Bigg(\frac{300\,{\rm GeV}}{1/R}\Bigg)^2\Bigg(\frac{30}{M_sR}\Bigg)^2\quad.
\label{5dst}
\eear
The current constraints on the magnitude of $S$ and $T$ from the precision measurements are roughly $0.2$.
With $c_2$ being of order unity, $S^{UV}$ is sufficiently small to be neglected in the total $S$ contributions.
However, $c_1$ of order unity would give a sizable $T^{UV}$ if the Higgs mass is much larger than $200\, {\rm GeV}$.
Thus, we could lose the predictability of $T$ in the region of large Higgs mass, even if the KK sum converges.
To extract reliable predictions from the KK sum alone, we may need to have a naturally smaller $c_1$ than of order unity.
We next argue that we indeed expect $c_1$ to be as small as 0.1. Then, $T^{UV}$ can be safely neglected in the range of Higgs mass
discussed in this paper.

The key observation is that $M_s\sim 30\cdot 1/R$ is the scale where 5D QCD coupling becomes nonperturbative, while
the electroweak sector remains perturbative and is still described by the effective 5D standard model.
Because the Higgs fields are QCD-neutral, couplings to the quark sector must be invoked to generate the custodial symmetry-violating operator (\ref{Toperator}).
The largest such coupling is the top Yukawa coupling.
The 5D top Yukawa coupling, $\hat{\lambda}_t$ has the mass dimension $-1/2$, and the dimensionless loop expansion parameter in 5D
is given by
\be
\frac{\hat{\lambda}_t^2 \,M_s}{24\,\pi^3}\,=\,\frac{(\pi R\lambda_t^2) M_s}{24\,\pi^3}\,\sim\,\frac{\pi R\,M_s}{24\,\pi^3}\,\sim\,\frac{30}{24\,\pi^2}
\,\sim\,0.13 \quad .
\ee
where $\lambda_t\,\sim\,1$ is the 4D top Yukawa coupling and we have used the relation $\hat{\lambda}_t=\sqrt{\pi R}\,\lambda_t$. The factor
$24\,\pi^3$ is from the 5D momentum integration.
This indicates that the top Yukawa coupling of the Higgs fields to the quarks is still perturbative at the scale $M_s$, and $c_1$ can be expected
to contain this factor.
At the scale where the electroweak sector becomes nonperturbative, which is somewhat higher than $M_s$,
additional contributions to (\ref{Toperator}) will be generated by strong electroweak dynamics, possibly without any approximate custodial symmetry,
but then the suppression scale is higher than $M_s$, which again makes $c_1\,\lae\,0.1$ . By contrast, there is
no obvious reason to expect $c_2$ to be smaller than of order unity.
With these estimates, $T^{UV}$ and $S^{UV}$ from (\ref{5dst}) are small enough to be neglected in calculating $S$ and $T$ contributions.

\begin{figure}[t]
\begin{center}
\scalebox{1}[1]{\includegraphics{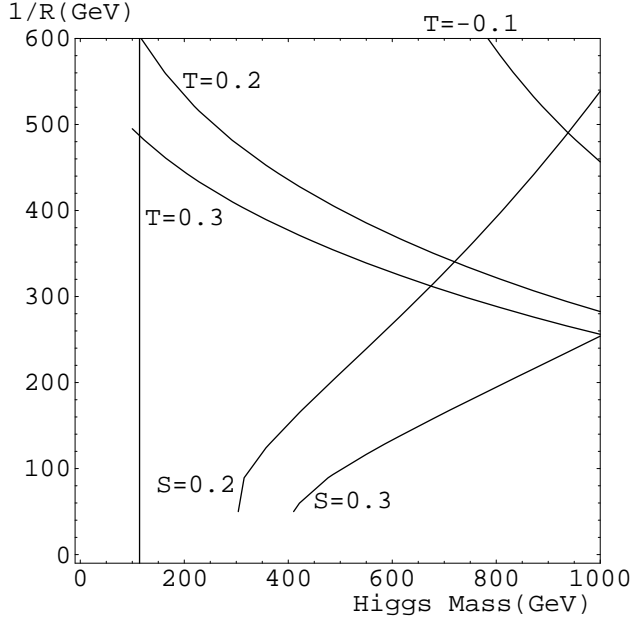}}
\par
\vskip-2.0cm{}
\end{center}
\caption{\small Some contours of total $S,T$ from the standard model and its higher KK modes in the 5D UED model on $S^1/Z_2$.
Here $m_{H}^{\rm ref}=115\,{\rm GeV}\,,\,m_t=174\,{\rm GeV}$. Up to 11 KK levels are included. The vertical line is the
direct search limit $m_H \ge 114\,{\rm GeV}$ (95\%C.L.) \cite{Heister:2001kr}.}
\label{contours}
\end{figure}
Having seen that the KK sum is reliable, we now analyze the consequences of the KK contributions to $S$ and $T$ by considering
the current combined $(S,T)$ constraints from the elecroweak precision measurements.
It is helpful first to see how the total $S$ and $T$ vary in the $(m_H,1/R)$ parameter space to get a rough idea of
how the constraints from $S$ and $T$ shape the allowed region in the $(m_H,1/R)$ parameter space.
In Fig.\ref{contours}, we show a contour plot of some values of total $S$ and $T$ contributions
from the standard model and its higher KK modes in the $(m_H,1/R)$ parameter space.
We also include the direct-search limit of $m_H \ge 114 \,{\rm GeV}$(95\% confidence level (C.L.)) \cite{Heister:2001kr}.
Because of a compensation between positive KK quark contribution and negative KK Higgs contribution to $T$, we see that
as $m_H$ increases, the lower bound on $1/R$ from $T$ is relaxed. 
For even larger $m_H$, large positive contributions to $S$ from the Higgs KK modes make the region excluded. 
When $1/R$ is larger than $\sim 450 \,{\rm GeV}$, the constraint that $T$ may not be large and negative sets an upper bound on $m_H$.
This can be understood from the fact that the Higgs sector gives negative contributions to $T$ as in the usual standard
model. 

\begin{figure}[t]
\begin{center}
\scalebox{1}[1]{\includegraphics{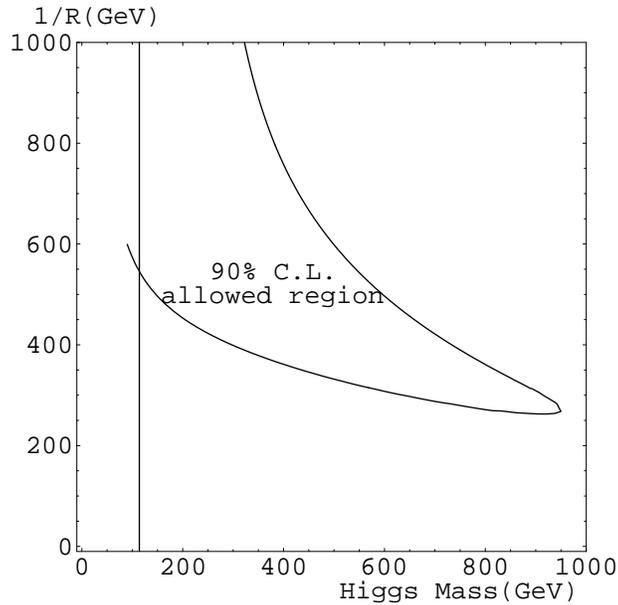}}
\par
\vskip-2.0cm{}
\end{center}
\caption{\small The 90\% C.L. allowed region in the 5D UED model on $S^1/Z_2$. Up to 11 KK levels are included.
Also shown is the direct search limit $m_H \ge 114\,{\rm GeV}$.}
\label{90percent}
\end{figure}
Because the constraints on the $S$ and $T$ parameters have a strong correlation \cite{Erler:ig}, separate $S$ and $T$ constraints
are incomplete. We therefore consider the current combined $(S,T)$ constraints to find the allowed region in the $(m_H,1/R)$ parameter space.
To find a (90\%) confidence level region, we analyze $\Delta \chi^2$ contours in the $(m_H,1/R)$ parameter space.
For this purpose, we may think of the $(m_H,1/R)$ parameters as a change of variables from $(S,T)$ because the number of
fitting parameters is two in both cases. Thus, we can simply use the $\Delta \chi^2$ contours
in the $(S,T)$ plane, for example, in Ref.\cite{Erler:ig}.
The resulting 90\% C.L. allowed region is shown in Fig.\ref{90percent}.
The region of smaller $1/R$ and larger $m_H$ than would be allowed from separate $S$ and $T$ constraints appears as a consequence
of the correlation between the $S$ and $T$ constraints.
The boundary of the region away from the tip is largely determined by $T$ constraints.
For $m_H\sim 800 \,{\rm GeV}$, even $1/R\sim 250\,{\rm GeV}$ is possible, and this should be testable
in the next collider experiments \cite{Cheng:2002ab,Rizzo:2001sd}.

\section{Two universal extra dimensions on $T^2/Z_2$} \setcounter{equation}{0}

In the case of one extra dimension, the KK contributions to $S,T$ and $U$ converge rapidly before
encountering the cutoff $M_s$, and the contributions from physics above $M_s$ are sufficiently small to be neglected.
Thus, practically the presence of $M_s$ is not significant. 
However, the KK sum diverges logarithmically in the 6D standard 
model, so we cannot
expect a reliable estimate from only summing the KK modes. A possible 
procedure is
to sum the KK modes up to the cutoff of the 6D model and then, as described 
in section 3,
to represent the physics beyond the cutoff by an appropriate operator. A 
problem with this
procedure is that while each term in the KK sum maintains 4D gauge invariance, the truncated sum  
is not expected
to respect the the full 6D gauge invariance upon which the 6D standard 
model is based \footnote{H.-U.Y. thanks Takemichi Okui for
discussions of this point at TASI 2002.}. As
noted below, however, the natural cutoff on the effective 6D theory is at 
about the fifth or sixth
KK level. With successive terms falling like $1/j$ and with the high energy 
contribution
represented by a 6D-gauge-invariant operator, we expect the lack of 6D gauge 
invariance to be relatively small - perhaps no more than a $20\%$ effect. We adopt this procedure 
with the
understanding that unlike the 5D case, only rough estimates are being 
provided in
six dimensions.

The most stringent estimate on $M_s$ in 6D comes from the naturalness
of the Higgs mass under quadratically divergent radiative corrections \cite{Appelquist:2002ft}. For a valid effective-theory description,
the six dimensional Higgs mass parameter $\hat{M}_H$ (the coefficient of the quadratic term of the 6D Higgs field)
should be below $M_s$, but at the same time, it shouldn't be small compared to the one-loop radiative correction on naturalness grounds:
\be
M_s \,\,>\,\,\hat{M}_H\,\,\gae\,\,\delta \hat{M}_H\,\sim\,\sqrt{\frac{\hat{\lambda}M_s^2}{128\pi^3}}\,M_s\quad,
\ee
where $\hat{\lambda}$ is the Higgs self coupling in 6D. The factor of $128\pi^3$ arises from the six dimensional momentum integral.
This gives the following relation involving the Higgs VEV $v=246\,{\rm GeV}$:
\be
v=\bigg[\pi\,R\,M_s\,(\hat{\lambda}M_s^2)^{-1/2}\bigg]\hat{M}_H\,\,\gae\,\,\frac{1}{\sqrt{128\pi}}(R\,M_s)^2\,R^{-1}\quad.
\ee
Taking $1/R$ of a few hundred ${\rm GeV}$ gives $R\,M_s\,\sim\,5$.
This result is similar to an estimate using the renormalization group analysis of both gauge couplings and Higgs self
coupling \cite{Arkani-Hamed:2000hv} showing that $M_s$ should be around five times of the compactification scale.
We therefore take $M_s\,\sim\,5/R$ in the following.

The contributions to $S$ and $T$ from physics above $M_s$, estimated in (\ref{deltaT}) and (\ref{deltaS}), can be written as
\bear
&&T^{UV}=c_1\cdot\,0.57\,\Bigg(\frac{m_H}{200\,{\rm GeV}}\Bigg)^2\Bigg(\frac{300\,{\rm GeV}}{1/R}\Bigg)^2\Bigg(\frac{5}{M_sR}\Bigg)^2\quad,\nonumber \\
&&S^{UV}=c_2\cdot\,0.17\,\Bigg(\frac{300\,{\rm GeV}}{1/R}\Bigg)^2\Bigg(\frac{5}{M_sR}\Bigg)^2\quad.
\label{6dst}
\eear
As in the 5D case, we expect $c_2$ to be a parameter of order unity. But in
contrast to 5D, there may be no good reason to anticipate that $c_1$ should
be less than unity. The reason is that in
6D, the scale at which the electroweak interactions (including the Yukawa
couplings to the top quark and other fermions) become strong is not much
above $M_s$, the scale at which 6D QCD becomes strong. Thus,
the breaking of custodial symmetry encoded in the operator (\ref{Toperator}) may 
be near-maximal. Since these estimates are crude, however, we
will allow in the estimates below for both maximal breaking of
custodial symmetry ($c_1 \approx 1$) as well as the presence of some
suppression of this breaking ($c_1 \approx 0.1$).

\begin{figure}[t]
\begin{center}
\scalebox{1}[1]{\includegraphics{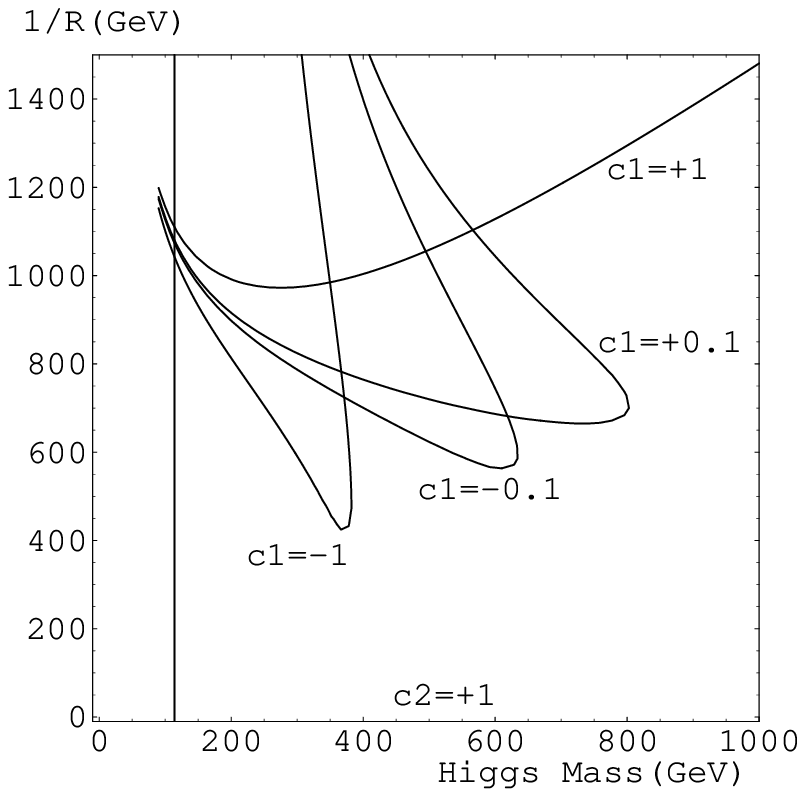} \includegraphics{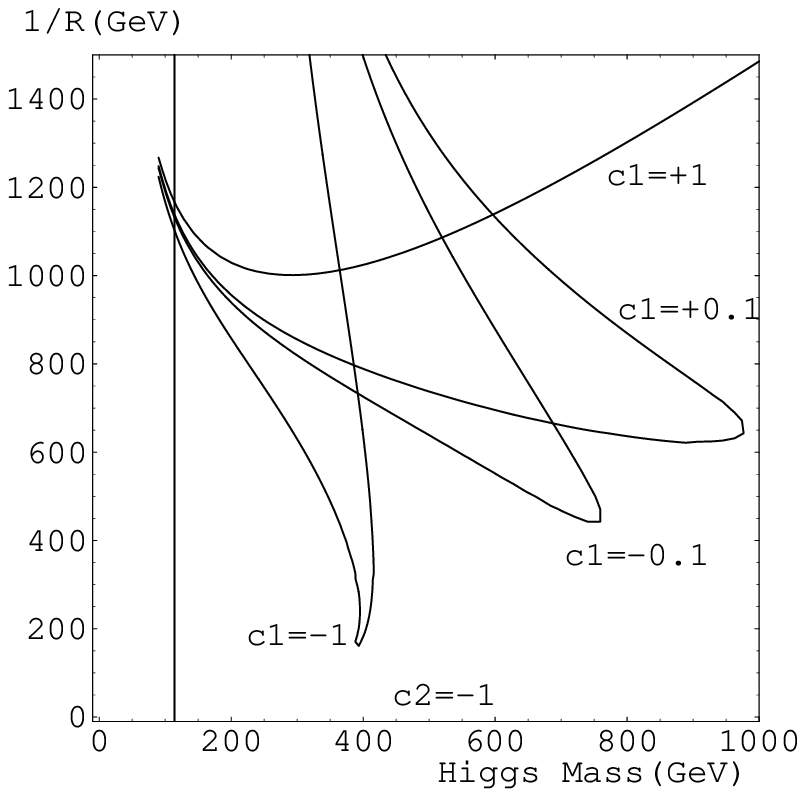}}
\par
\vskip-2.0cm{}
\end{center}
\caption{\small The 90\% C.L. allowed regions for several values of $c_1$ and $c_2$ in the 6D UED model on $T^2/Z_2$.
Also shown is the direct search limit $m_H \ge 114\,{\rm GeV}$.}
\label{6D90percent}
\end{figure}
In Fig.\ref{6D90percent}, we show several 90\% C.L. allowed regions
taking $c_1=\pm \,1$ or $\pm \,0.1$, and $c_2=\pm \,1$.
As mentioned above, the contribution from physics below $M_s$ is estimated by summing KK contributions up to $M_s\,=5/R$.
The plot shows very different characteristic features for different signs and magnitudes of $c_1$ and $c_2$.
It should be taken only to indicate possibilities, though, because of the uncertainty in the estimates of $c_1$ and $c_2$.

First, consider the case $c_1=\,\pm 0.1$. In this case, the contributions to $T$ from physics above $M_s$ do not affect the shapes of the regions significantly.
However, an important dependence on the sign of $c_2$
appears.
For negative $c_2$ (the right figure), a region of larger Higgs mass can be allowed, compared to the case of
positive $c_2$ (the left figure). This can be understood from the fact that this region is constrained by large positive
KK contributions to $S$ as can be seen in Fig.\ref{contours}. With negative $c_2$, the contribution from physics above $M_s$ can cancel the KK
contributions in this region, relieving the constraint from $S$.
Since no such cancellation is involved when $c_2=\,1$, 
the left figure may describe a more generic allowed region for the case $c_1=\,\pm 0.1$.

In the (perhaps more likely) case $c_1=\,\pm 1$, the contributions to $T$ from physics 
above the cutoff have significant effects on the shape of the regions,
while the contributions to $S$ from above the cutoff play a lesser role in determining the allowed regions. When $c_1=\,+1$,
a region where both $m_H$ and $1/R$ are large can be allowed, because the large negative total
contributions to $T$ from KK modes with large $m_H$ can be compensated by the positive UV contribution to $T$.
With $c_1=\,-1$, Higgs masses lighter than $\sim 400\,{\rm GeV}$ are preferred. 
This is because
the negative UV contribution to $T$ can then be cancelled by the dominant positive contributions to $T$ from the KK top-quark doublets.
Higgs masses heavier than $\sim 400 \,{\rm GeV}$ are excluded in the $c_1=\,-1$ case because they too give a negative total KK contribution to $T$.

Although we can't extract precise information from the plots of Fig.\ref{6D90percent}, due to the uncertainty in the signs and magnitudes of the coefficients $c_1$ and $c_2$,
the above results do tell us that the possibility of a large Higgs mass and a relatively small compactification scale is not excluded in 6D UED theories.

\section{Conclusions} \setcounter{equation}{0}

The discovery of additional spatial dimensions accessible to the
standard-model
fields (universal extra dimensions) would be a spectacular realization of physics beyond the standard
model.
The first study \cite{Appelquist:2000nn} of the constraint on the compactification scale, $1/R$, from precision electroweak measurements
in theories of universal extra dimensions
gave the bound $1/R\,\gae\,300\,{\rm GeV}$. But an assumption of that analysis was that the Higgs mass $m_H$ is less than $250\,{\rm GeV}$.
In this paper, we have considered the precision electroweak constraints in terms of the $S$ and $T$ parameters
without assuming that $m_H \lae250\,{\rm GeV}$.


We have shown that current precision measurements, when analyzed
with both the compactification scale $1/R$ and the Higgs mass $m_H$ taken
to be free parameters, lead to a lower bound on $1/R$ that is quite
sensitive to
$m_H$ and can be as low as $~ 250 \,{\rm GeV}$. This becomes possible if $m_H$
is larger than allowed in the minimal standard model -- as large as $~800 \,
{\rm GeV}$.
Equivalently, in the presence of low-scale universal extra dimensions,
precision
measurements allow a considerably larger $m_H$ than in the framework of the
minimal standard model. The main reason for this is that the negative contributions to the $T$ parameter from the Higgs boson and its KK partners
can be cancelled by the positive contributions from KK top quarks.

A light compactification scale would have important
consequences for the possibility of direct detection of KK particles in the next collider experiments \cite{Cheng:2002ab,Rizzo:2001sd,Macesanu:2002ew}. 
The KK dark matter density \cite{Servant:2002aq}
and its direct or indirect detection \cite{Cheng:2002ej,Hooper:2002gs,Servant:2002hb,Majumdar:2002mw} are
sensitive both
to the compactification scale and to the Higgs mass through the rates of
the Higgs-mediated
processes. It would be interesting to reanalyze them in the allowed
$(m_H,1/R)$ parameter
region obtained in this paper.

\bigskip
 {\bf Acknowledgements:} \
 We would like to thank Bogdan Dobrescu for many critical comments in the early stages of this work, and Tatsu Takeuchi for an important 
 discussion on section 2.2.
 We also thank Hsin-Chia Cheng and Eduardo Ponton for helpful discussions. This work was supported by DOE under contract DE-FG02-92ER-40704.

\section*{Appendix : Summary of one-loop KK contributions to gauge-boson self energies}
\renewcommand{\theequation}{A.\arabic{equation}}
\setcounter{equation}{0}

In this appendix, we summarize the calculation of one-loop diagrams with intermediate KK particles for the (zero mode) gauge-boson propagators.
We introduce the higher dimensional analog of the $R_{\xi}$ gauge with $\xi=1$, in which extra dimensional components of gauge
bosons can be treated as 4D scalar fields without any mixed kinetic terms with 4D components. Because KK number is conserved at vertices and
the external lines are zero modes, all KK particles in one-loop diagrams are in the same $j$th level.
We group the diagrams into five classes such that quadratic divergences cancel within a class.\\\\
(a) Loops with KK quarks of the third generation\\
(b) Loops with KK gauge bosons, in which at least one internal line is a 4D component and loops with KK ghosts\\
(c) Loops with KK gauge bosons with extra dimensional components (should be multiplied by $\delta$, the number of extra dimensions)\\
(d) Loops with KK particles from the Higgs sector\\
(e) Loops with one KK gauge boson and one KK particle from the Higgs sector\\\\
In the following, $s_w \equiv \sin\theta_w$, $c_w \equiv \cos\theta_w$, $E \equiv \frac{2}{\varepsilon}-\gamma+\log4\pi$, and
\bear
\Delta^2_j(m_1^2,m_2^2,x) \equiv M_j^2-x(1-x)p^2+(1-x)m_1^2+x\,m_2^2 
\eear
where $M_j^2 \equiv \big(\frac{j}{R}\big)^2$.

{\bf (1) $WW$ self energy}
\bear
\Pi^{j(a)}_{WW}(p^2)&=&\frac{\alpha}{4\pi}\frac{-6}{s_w^2}\int^1_0 dx\,\Big(E-\log\Delta^2_j(0,m_t^2,x)\Big)\Big(2x(1-x)p^2-x\,m_t^2\Big)\nonumber\\
\Pi^{j(b)}_{WW}(p^2)&=&\frac{\alpha}{4\pi}\frac{c_w^2}{s_w^2}\int^1_0 dx\,
\Big(E-\log\Delta^2_j(m_W^2,m_Z^2,x)\Big) \nonumber \\
&&\qquad\qquad\qquad\cdot\Big(2(-4x^2+4x+1)p^2+(3-4x)m_Z^2+(4x-1)m_W^2\Big)\nonumber \\
&+&\frac{\alpha}{4\pi}\int^1_0 dx\,\Big(E-\log\Delta^2_j(m_W^2,0,x)\Big)\Big(2(-4x^2+4x+1)p^2+(4x-1)m_W^2\Big)\nonumber \\
\Pi^{j(c)}_{WW}(p^2)&=&\frac{\alpha}{4\pi}\frac{c_w^2}{s_w^2}\int^1_0 dx\,
\Big(E-\log\Delta^2_j(m_W^2,m_Z^2,x)\Big) \nonumber \\
&&\qquad\qquad\qquad\cdot\Big(-(4x^2-4x+1)p^2+(1-2x)m_Z^2+(2x-1)m_W^2\Big)\nonumber \\
&+&\frac{\alpha}{4\pi}\int^1_0 dx\,\Big(E-\log\Delta^2_j(m_W^2,0,x)\Big)\Big(-(4x^2-4x+1)p^2+(2x-1)m_W^2\Big)\nonumber \\
\Pi^{j(d)}_{WW}(p^2)&=&\frac{\alpha}{4\pi}\frac{1}{4s_w^2}\int^1_0 dx\,
\Big(E-\log\Delta^2_j(m_W^2,m_Z^2,x)\Big) \nonumber \\
&&\qquad\qquad\qquad\cdot\Big(-(4x^2-4x+1)p^2+(1-2x)m_Z^2+(2x-1)m_W^2\Big)\nonumber \\
&+&\frac{\alpha}{4\pi}\frac{1}{4s_w^2}\int^1_0 dx\,\Big(E-\log\Delta^2_j(m_W^2,m_H^2,x)\Big) \nonumber \\
&&\qquad\qquad\qquad\cdot\Big(-(4x^2-4x+1)p^2+(1-2x)m_H^2+(2x-1)m_W^2\Big)\nonumber \\
\Pi^{j(e)}_{WW}(p^2)&=&\frac{\alpha}{4\pi}\frac{-1}{s_w^2}\int^1_0 dx\,\Big(E-\log\Delta^2_j(m_W^2,m_H^2,x)\Big)m_W^2 \nonumber \\
&+&\frac{\alpha}{4\pi}(-s_w^2)\int^1_0 dx\,\Big(E-\log\Delta^2_j(m_W^2,m_Z^2,x)\Big)m_Z^2 \nonumber \\
&+&\frac{\alpha}{4\pi}(-1)\int^1_0 dx\,\Big(E-\log\Delta^2_j(m_W^2,0,x)\Big)m_W^2 \nonumber
\eear

{\bf (2) $ZZ$ self energy}
\bear
\Pi^{j(a)}_{ZZ}(p^2)&=&\frac{\alpha}{4\pi}\frac{-3 + 8 s_w^2-\frac{32}{3}s_w^4}{s_w^2c_w^2}
\int^1_0 dx\,\Big(E-\log\Delta^2_j(m_t^2,m_t^2,x)\Big)\Big(2x(1-x)p^2\Big)\nonumber\\
&+&\frac{\alpha}{4\pi}\frac{3}{s_w^2c_w^2}\int^1_0 dx\,\Big(E-\log\Delta^2_j(m_t^2,m_t^2,x)\Big)m_t^2\nonumber \\
&+&\frac{\alpha}{4\pi}\frac{-3 + 4 s_w^2-\frac{8}{3}s_w^4}{s_w^2c_w^2}
\int^1_0 dx\,\Big(E-\log\Delta^2_j(0,0,x)\Big)\Big(2x(1-x)p^2\Big)\nonumber\\
\Pi^{j(b)}_{ZZ}(p^2)&=&\frac{\alpha}{4\pi}\frac{c_w^2}{s_w^2}\int^1_0 dx\,\Big(E-\log\Delta^2_j(m_W^2,m_W^2,x)\Big)\Big((-8x^2+14x-1)p^2+2m_W^2\Big)\nonumber\\
\Pi^{j(c)}_{ZZ}(p^2)&=&\frac{\alpha}{4\pi}\frac{2c_w^2}{s_w^2}\int^1_0 dx\,\Big(E-\log\Delta^2_j(m_W^2,m_W^2,x)\Big)\Big((-2x^2+3x-1)p^2\Big)\nonumber\\
\Pi^{j(d)}_{ZZ}(p^2)&=&\frac{\alpha}{4\pi}\frac{1}{4s_w^2c_w^2}\int^1_0 dx\,\Big(E-\log\Delta^2_j(m_Z^2,m_H^2,x)\Big)\nonumber \\
&&\qquad\qquad\qquad\cdot\Big((-4x^2+4x-1)p^2+(1-2x)m_H^2+(2x-1)m_Z^2\Big)\nonumber \\
&+&\frac{\alpha}{4\pi}\frac{(c_w^2-s_w^2)^2}{2s_w^2c_w^2}\int^1_0 dx\,\Big(E-\log\Delta^2_j(m_W^2,m_W^2,x)\Big)\Big((-2x^2+3x-1)p^2\Big)\nonumber\\
\Pi^{j(e)}_{ZZ}(p^2)&=&\frac{\alpha}{4\pi}\frac{-1}{s_w^2c_w^2}\int^1_0 dx\,\Big(E-\log\Delta^2_j(m_Z^2,m_H^2,x)\Big)m_Z^2 \nonumber \\
&+&\frac{\alpha}{4\pi}(-2s_w^2)\int^1_0 dx\,\Big(E-\log\Delta^2_j(m_W^2,m_W^2,x)\Big)m_Z^2 \nonumber
\eear

{\bf (3) $Z\gamma$ self energy}
\bear
\Pi^{j(a)}_{Z\gamma}(p^2)&=&\frac{\alpha}{4\pi}\frac{-4+\frac{32}{3}s_w^2}{s_wc_w}
\int^1_0 dx\,\Big(E-\log\Delta^2_j(m_t^2,m_t^2,x)\Big)\Big(2x(1-x)p^2\Big)\nonumber\\
&+&\frac{\alpha}{4\pi}\frac{-2+\frac{8}{3}s_w^2}{s_wc_w}
\int^1_0 dx\,\Big(E-\log\Delta^2_j(0,0,x)\Big)\Big(2x(1-x)p^2\Big)\nonumber\\
\Pi^{j(b)}_{Z\gamma}(p^2)&=&\frac{\alpha}{4\pi}\frac{c_w}{s_w}
\int^1_0 dx\,\Big(E-\log\Delta^2_j(m_W^2,m_W^2,x)\Big)\Big((-8x^2+14x-1)p^2+2m_W^2\Big)\nonumber\\
\Pi^{j(c)}_{Z\gamma}(p^2)&=&\frac{\alpha}{4\pi}\frac{2c_w}{s_w}
\int^1_0 dx\,\Big(E-\log\Delta^2_j(m_W^2,m_W^2,x)\Big)\Big((-2x^2+3x-1)p^2\Big)\nonumber\\
\Pi^{j(d)}_{Z\gamma}(p^2)&=&\frac{\alpha}{4\pi}\frac{c_w^2-s_w^2}{s_wc_w}
\int^1_0 dx\,\Big(E-\log\Delta^2_j(m_W^2,m_W^2,x)\Big)\Big((-2x^2+3x-1)p^2\Big)\nonumber\\
\Pi^{j(e)}_{Z\gamma}(p^2)&=&\frac{\alpha}{4\pi}\frac{2s_w}{c_w}
\int^1_0 dx\,\Big(E-\log\Delta^2_j(m_W^2,m_W^2,x)\Big)m_W^2\nonumber
\eear

{\bf (4) $\gamma\gamma$ self energy}
\bear
\Pi^{j(a)}_{\gamma\gamma}(p^2)&=&\frac{\alpha}{4\pi}\frac{-32}{3}
\int^1_0 dx\,\Big(E-\log\Delta^2_j(m_t^2,m_t^2,x)\Big)\Big(2x(1-x)p^2\Big)\nonumber\\
&+&\frac{\alpha}{4\pi}\frac{-8}{3}
\int^1_0 dx\,\Big(E-\log\Delta^2_j(0,0,x)\Big)\Big(2x(1-x)p^2\Big)\nonumber\\
\Pi^{j(b)}_{\gamma\gamma}(p^2)&=&\frac{\alpha}{4\pi}
\int^1_0 dx\,\Big(E-\log\Delta^2_j(m_W^2,m_W^2,x)\Big)\Big((-8x^2+14x-1)p^2+2m_W^2\Big)\nonumber\\
\Pi^{j(c)}_{\gamma\gamma}(p^2)&=&\frac{\alpha}{4\pi}
\int^1_0 dx\,\Big(E-\log\Delta^2_j(m_W^2,m_W^2,x)\Big)\Big(2(-2x^2+3x-1)p^2\Big)\nonumber\\
\Pi^{j(d)}_{\gamma\gamma}(p^2)&=&\frac{\alpha}{4\pi}
\int^1_0 dx\,\Big(E-\log\Delta^2_j(m_W^2,m_W^2,x)\Big)\Big(2(-2x^2+3x-1)p^2\Big)\nonumber\\
\Pi^{j(e)}_{\gamma\gamma}(p^2)&=&\frac{\alpha}{4\pi}
\int^1_0 dx\,\Big(E-\log\Delta^2_j(m_W^2,m_W^2,x)\Big)\big(-2m_W^2\big)\nonumber
\eear

 \vfil \end{document}